\documentclass[12pt,preprint]{aastex}

\begin{document}

\title{Color, Structure, and Star Formation History of Dwarf Galaxies over the
last $\sim3$ Gyr with GEMS and SDSS}

\author{Fabio D. Barazza, Shardha Jogee}
\affil{Department of Astronomy, University of Texas at Austin, 1 University
Station C1400, Austin, TX 78712-0259, USA}
\email{barazza@astro.as.utexas.edu, sj@astro.as.utexas.edu}

\author{Hans-Walter Rix, Marco Barden, Eric F. Bell}
\affil{Max-Planck Institute for Astronomy, Koenigstuhl 17, D-69117 Heidelberg,
Germany}

\author{John A. R. Caldwell}
\affil{McDonald Observatory, University of Texas, Fort Davis, TX 79734, USA}

\author{Daniel H. McIntosh}
\affil{Department of Astronomy, University of Massachusetts, Amherst, MA 01003,
U.S.A}

\author{Klaus Meisenheimer}
\affil{Max-Planck Institute for Astronomy, Koenigstuhl 17, D-69117 Heidelberg,
Germany}

\author{Chien Y. Peng}
\affil{Space Telescope Science Institute, 3700 San Martin Dr., Baltimore, MD
21218, USA}

\and

\author{Christian Wolf}
\affil{Department of Physics, Denys Wilkinson Bldg., University of Oxford,
Keble Road, Oxford OX1 3RH, U.K.}

\begin{abstract}
We present a study of the colors, structural properties, and star formation
histories for a sample of $\sim1600$ dwarfs over look-back times of $\sim3$ Gyr
($z=0.002-0.25$). The sample consists of 401 distant dwarfs drawn from the
Galaxy Evolution from Morphologies and SEDs (GEMS) survey, which provides high
resolution {\it Hubble Space Telescope (HST)} Advanced Camera for Surveys (ACS)
images and accurate redshifts, and of 1291 dwarfs at 10--90 Mpc compiled from
the Sloan Digitized Sky Survey (SDSS). The sample is complete down to an
effective surface brightness of 22 mag arcsec$^{-2}$ in $z$ and includes dwarfs
with $M_g=-18.5$ to $-14$ mag. Rest-frame luminosities in Johnson $UBV$ and
SDSS $ugr$ filters are provided by the COMBO-17 survey and structural
parameters have been determined by S\'ersic fits. We find that the GEMS dwarfs
are bluer than the SDSS dwarfs by $\sim0.13$ mag in $g-r$, which is consistent
with the color evolution over $\sim2$ Gyr of star formation histories involving
moderate starbursts and long periods of continuous star formation. The full
color range of the samples cannot be reproduced by single starbursts of
different masses or long periods of continuous star formation alone.
Furthermore, an estimate of the mechanical luminosities needed for the gas in
the GEMS dwarfs to be completely removed from the galaxies shows that a
significant number of low luminosity dwarfs are susceptible to such a complete
gas loss, {\it if} they would experience a starburst. On the other hand, a
large fraction of more luminous dwarfs is likely to retain their gas. We also
estimate the star formation rates per unit area for the GEMS dwarfs and find
good agreement with the values for local dwarfs.
\end{abstract}


\section{Introduction}
Until recently the study of dwarf galaxies has been primarily concentrated
either on clusters, where a large number of dwarfs can be observed within a
relatively small area of sky \citep[e.g.,][]{bin91,tre98,dri01}, or on the
Local Group, where dwarfs can be studied in great detail due to their proximity
\citep[for recent reviews see][]{mat98,ber00,gre00}. Dwarfs outside of these
regimes are numerous, but widespread. Many of them are associated with giant
galaxies, forming galaxy groups \citep{kar05}, which to some extent have been
studied photometrically \citep{bre98,bre99,bre00,jer00} and kinematically
\citep{bot90,kar99}. Studies of dwarfs between clusters and groups are rather
rare and are mainly composed of small samples, which have been selected more or
less randomly \citep{mak99,bar01,par02} or examine specific types of dwarfs,
e.g., blue compact dwarfs \citep[BCDs;][]{noe03,gil03}.

The evolution of dwarfs is a complex problem, where evolutionary paths may
depend on a variety of external and internal factors. Our knowledge of the
local volume ($<8$ Mpc) has deepened, in particular due to strong efforts in
determining distances to many nearby galaxies \citep[and references
therein]{kar03}. However, it is still unclear what governs the evolution of
dwarfs in low density regions and how the different morphological types form.
Progress has in part been hampered by the fact that many early studies are
based on small samples, suffer from small number statistics and had systematic
selection biases. Headway toward constraining the evolutionary history of dwarf
galaxies and identifying the fundamental physical processes involved requires,
as a first step, the study of large, statistically significant samples of
dwarfs, selected over a large region of the sky without systematic biases.

In a cosmological context, empirical constraints on dwarf galaxies at different
look-back times are essential for testing hierarchical models of galaxy
formation \citep{whi91,ste02}. One such aspect is the possibility that star
formation in low mass dark matter halos gets suppressed due to a number of
reasons, including the ejection of gas out of a shallow potential well by
early, possibly primordial, episodes of star formation \citep{sil03,bur04}.

Many questions remain unresolved in simulations. Over what epochs does the
suppression happen: is this a primordial process only, or do dwarfs present a
few Gyrs ago also exhibit signs of blowout? What types of masses and star
formation histories do dwarf galaxies exhibit {\it empirically}, and with such
empirically established star formation histories, are they expected to retain
their gas? What type of descendants would dwarfs at these epochs yield, if star
formation was turned off?

In this paper, we present a study of the properties of dwarf galaxies over the
last 3 Gyr ($z=0.002-0.25$)\footnote{We assume in this  paper a flat cosmology
with $\Omega_M = 1 - \Omega_{\Lambda} = 0.3$ and
$H_{\rm 0}$=70~km~s$^{-1}$~Mpc$^{-1}$.} drawn from GEMS \citep[Galaxy Evolution
from Morphologies and SEDs,][]{rix04} and the Sloan Digitized Sky Survey
\citep[SDSS,][]{aba04}. Our initial sample consists of 988 dwarfs from the GEMS
survey in the redshift range $z\sim0.09-0.25$ (corresponding to look-back times
$T_{back}$ of 1 to 3 Gyr), and a comparison local sample of 2847 dwarfs with
$z<0.02$ ($T_{back}\approx0.3$ Gyr) from SDSS. In this paper, we concentrate on
the colors, star formation rates (SFR), star formation histories, and
structural properties of dwarf galaxies. In a follow-up paper (Barazza et al.
in prep.) we will present a detailed morphological analysis of the dwarf sample
from GEMS and also take into account their environment. Throughout the paper we
will refer to the dwarf samples from GEMS and SDSS just as $dwarfs$ without
specifying classes (e.g., dE, Im etc.). However, we can assume that most of
them are Sm, Im, or BCD in the classification scheme of \citet{san84}, which
was also confirmed by a rough visual inspection. This assumptions is based on
the fact that both surveys particularly cover low density environments, where
late type dwarfs dominate \citep{kar04}. The rest of the paper is organized as
follows: in $\S$ \ref{sur} we give a brief description of the GEMS survey and
explain the sample selections; in $\S$ \ref{res} follows the presentation of
the comparisons performed and the resultant implications as well as a
discussion of the findings; finally the summary and conclusions are given in
$\S$ \ref{sum}.

\section{Observations, Sample Selection, and Analysis}\label{sur}
\subsection{GEMS survey}
GEMS is a two-band (F606W and F850LP) {\it Hubble Space Telescope} ($HST$)
large-area (800 arcmin$^2$) survey using the Advanced Camera for Surveys (ACS),
with accurate redshifts from Combo-17 \citep{wol04}. The principal aim of the
GEMS survey is the study of the evolution of galaxies out to $T_{back}\sim8$
Gyr ($z\sim1.0$), using morphologies and structural parameters as well as
spectral energy distributions (SEDs). The morphological information is provided
by a large-area (800 arcmin$^2$) two-filter ($V$ and $z$) imaging survey with
ACS on $HST$. The GEMS field is centered on the Chandra Deep Field-South and
reaches depths of $=28.3~(5\sigma)$ AB mag in F606W and $=27.1~(5\sigma)$ AB
mag in F850LP for compact sources. The actual GEMS survey provides
high-resolution ($\sim 0\farcs 05$ corresponding to 165 pc at $z\sim 0.2$) ACS
images for $\sim 8300$ galaxies with accurate redshifts
($\delta_z /(1+z)\sim0.02$ down to $R_{Vega}<23$ mag)\footnote{Henceforth, all
magnitudes are given in the VEGA system}. Details of the data reduction and
galaxy catalog construction are given in Caldwell et al. 2006, in prep.

The dwarf sample drawn from the GEMS survey (see $\S$ \ref{gsa}) has a median
redshift of $z=0.15$. Therefore, the structural parameters from the GEMS $z$
images correspond to a rest-frame filter between SDSS $i$ and SDSS $z$. This
has to be kept in mind and is further addressed in the discussion of the
results presented below. On the other hand, the COMBO-17 survey provides
absolute magnitudes in rest-frames $UBV$ and SDSS $ugr$, and can be compared
directly with the local dwarfs.

\subsection{The GEMS dwarf sample}\label{gsa}
We identified and extracted dwarfs from the GEMS survey by applying an absolute
magnitude cut of $M_g>-18.5$ mag. This limit is motivated by two findings.
First the luminosity functions of different clusters exhibit small dips around
$M_B\approx-18$ mag \citep[e.g.][]{tre02,mob03,pra04}, which might indicate the
transition between giants and dwarfs. Secondly, the typical luminosity, which
separates Sd from Sm galaxies is around $M_B=-18$ mag \citep{san85} and can be
considered as the transition luminosity between disk galaxies massive enough to
form spiral arms and disks with more irregular structures, commonly referred to
as dwarfs. Thus, the applied cut also has morphological implications. The GEMS
sample is complete down to $R=24$ mag, which roughly corresponds to a surface
brightness in $z$ of $\mu\approx23.5$ mag arcsec$^{-2}$. However, we limit our
GEMS dwarf sample to a effective surface brightness ($\mu_e$) in $z$ of
$\mu_e<22$ mag arcsec$^{-2}$, since this is the completeness limit of the SDSS
sample (see $\S$ \ref{sdsssa}). We correct for the effect of cosmological
surface brightness dimming using the factor $(1+z)^4$. Finally, we select the
redshift range $z=0.09-0.25$ for the GEMS dwarf sample. The lower limit
corresponds to three times the expected error in redshift for our dwarf sample
($\delta_z /(1+z)\approx 0.03$), which is slightly larger than the average
error of the GEMS sample, due to the lower luminosity of the dwarfs. The
selection of the upper redshift limit was guided by the goal to target dwarfs
over a large range of look-back times and potentially different evolutionary
stages, while ensuring that the sample at the higher redshift end is not
dominated by only bright galaxies, but includes a fair number of dwarfs in the
range $M_g=-14$ to $-18$ mag. These selection conditions yield a sample of 401
objects.

\begin{figure}
\epsscale{1}
\plotone{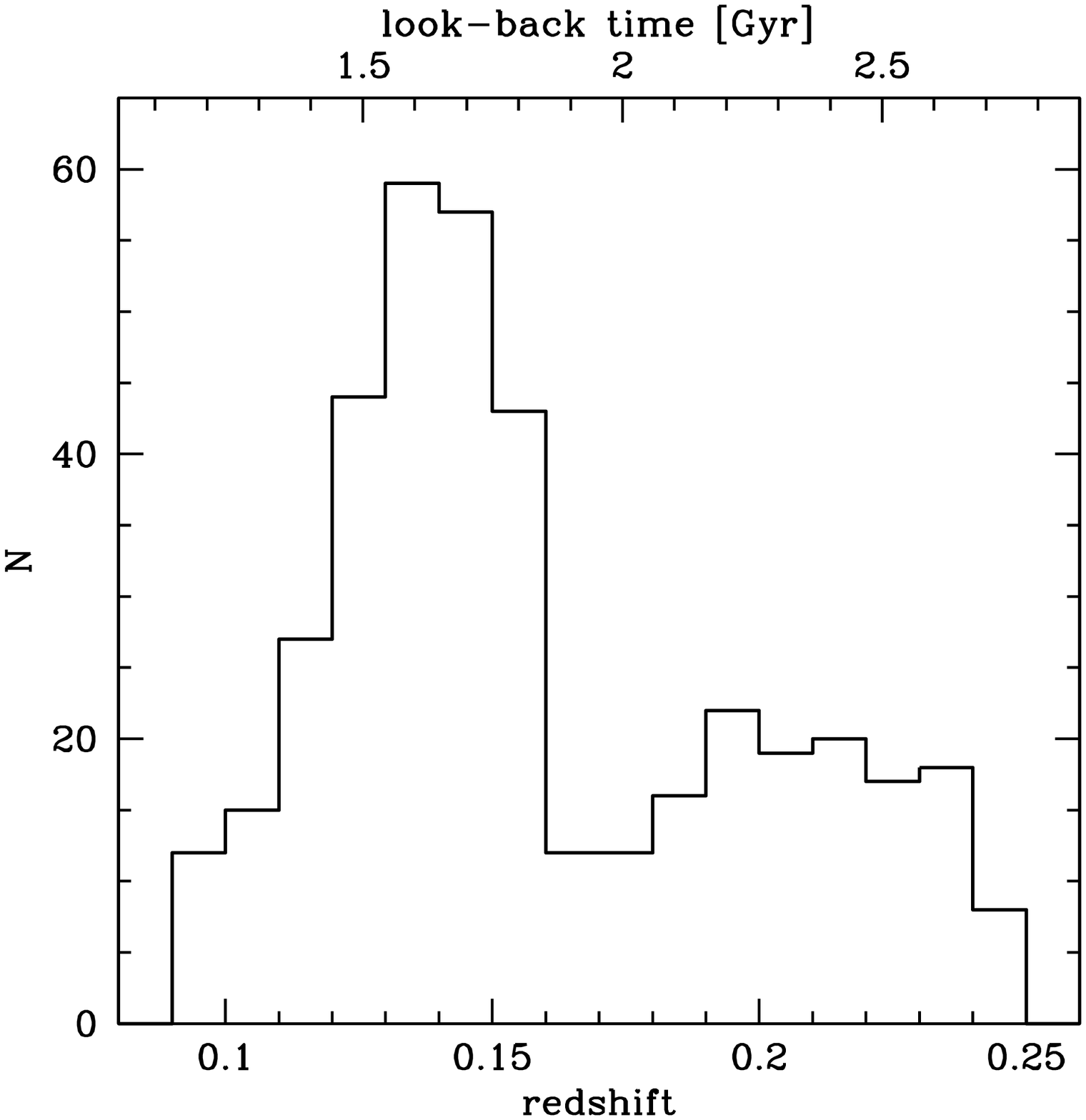}
\caption{The number of objects per redshift bin of our final GEMS sample (401
objects). The upper axis shows the approximate look-back time. The median
look-back time of the sample is $\sim1.8$ Gyr. The inhomogeneous distribution
is caused by cosmic variance.\label{red}}
\end{figure}

Our goal is to compare the GEMS sample to a sample from the NYU Value-Added
low-redshift Galaxy Catalog \citep[NYU-VAGC,][]{bla05} of the SDSS, whose
completeness limits are defined with respect to effective surface brightness.
We therefore fitted the GEMS dwarfs with a S\'ersic model \citep{ser68} using
GALFIT \citep{pen02}, which provides half-light radii and S\'ersic indices for
all objects. The fits were performed on the $z$-band images and we subsequently
limited our final sample to $\mu_e\leq22$ mag arcsec$^{-2}$. In view of the
completeness limit of the GEMS survey (see above), our final GEMS sample is
complete down to $\mu_e=22$ mag arcsec$^{-2}$ in $z$ for objects with $z<0.25$.
The redshift distribution of the final sample, which consists of 401 objects,
is shown in Figure \ref{red}. We note that the volume covered by this sample is
rather small ($\sim17000$ Mpc$^3$) and the sample may suffer from cosmic
variance and not be fully representative of the whole dwarf galaxy population
at look-back times around 2 Gyr. Nonetheless, the advantage of this sample is
that it allows us to study {\it HST}-based structural parameters of a complete
sample of dwarfs at earlier look-back times ($\sim3$ Gyr) than has been
possible to date.

\subsection{SDSS low redshift dwarf sample}\label{sdsssa}
\begin{figure}
\epsscale{1}
\plotone{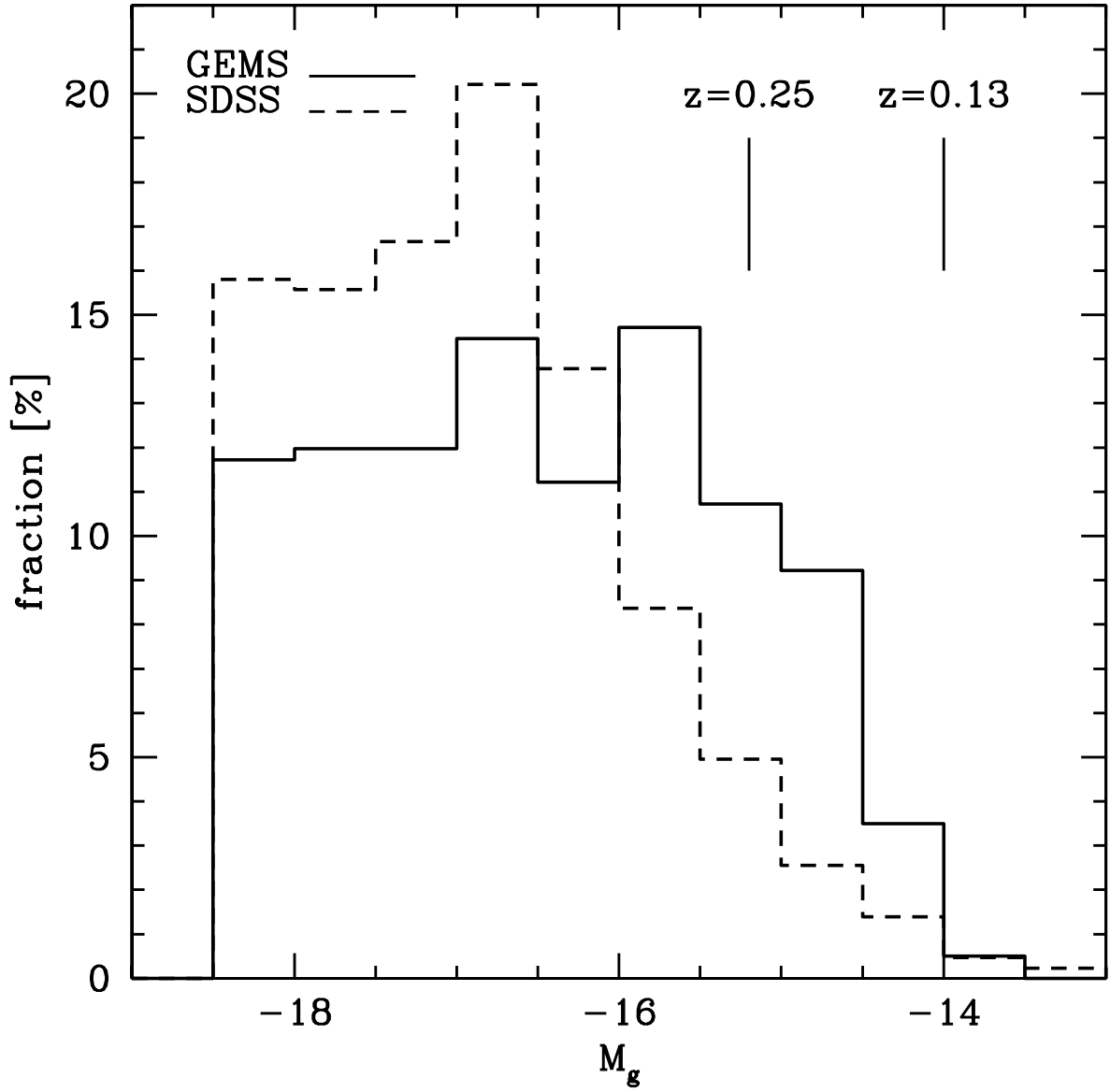}
\caption{The $g$-band luminosity distribution of the GEMS and SDSS samples. The
median value for the GEMS dwarf sample is $-16.51$ mag and for SDSS $-16.95$
mag. The completeness limits of the GEMS sample at two redshifts are
indicated.\label{lum}}
\end{figure}

SDSS is acquiring {\it ugriz} CCD imaging of $10^4$ deg$^2$ of the northern
Galactic sky and selecting $10^6$ targets for spectroscopy, most of them
galaxies with $r<17.77$ mag \citep{aba04}. The local dwarf sample is drawn from
the low-redshift catalog of the NYU-VAGC \citep{bla05}, which is based on the
second data release of SDSS.\footnote{The used data are therefore not affected
by a recently described error in model magnitudes of extended objects
\citep{str05}}. This low-redshift catalog consists of 28089 galaxies with
distances of 10--200 Mpc ($0.0033<z<0.05$), which have been determined by
correcting for peculiar velocities. The catalog provides, among other
properties, rest-frame absolute magnitudes as well as photometric parameters
from a S\'ersic fit in the SDSS $ugriz$ filters. The NYU-VAGC SDSS sample of
dwarfs was selected using the limits $M_g>-18.5$ mag and $z<0.02$. Finally, we
again applied a cut at an  effective surface brightness of 22 mag arcsec$^{-2}$
in $z$, which corresponds to the completeness limit of the NYU-VAGC, given with
respect to the $r$-band \citep{bla05}. The final sample of local dwarfs
consists of 1291 objects and is complete down to our surface brightness limit.
The luminosity distributions of the GEMS and SDSS samples are shown in Figure
\ref{lum}. The distributions are quite different in the sense that the fraction
of low luminosity dwarfs ($M_g>-16$ mag) is larger in the GEMS sample. On the
other hand, there is no large difference between the median values, which are
$-16.51$ mag and $-16.95$ mag for the GEMS and SDSS samples, respectively.

\section{Results and Discussion}\label{res}
\subsection{Global colors}\label{gcol}
\begin{figure}
\epsscale{1}
\plottwo{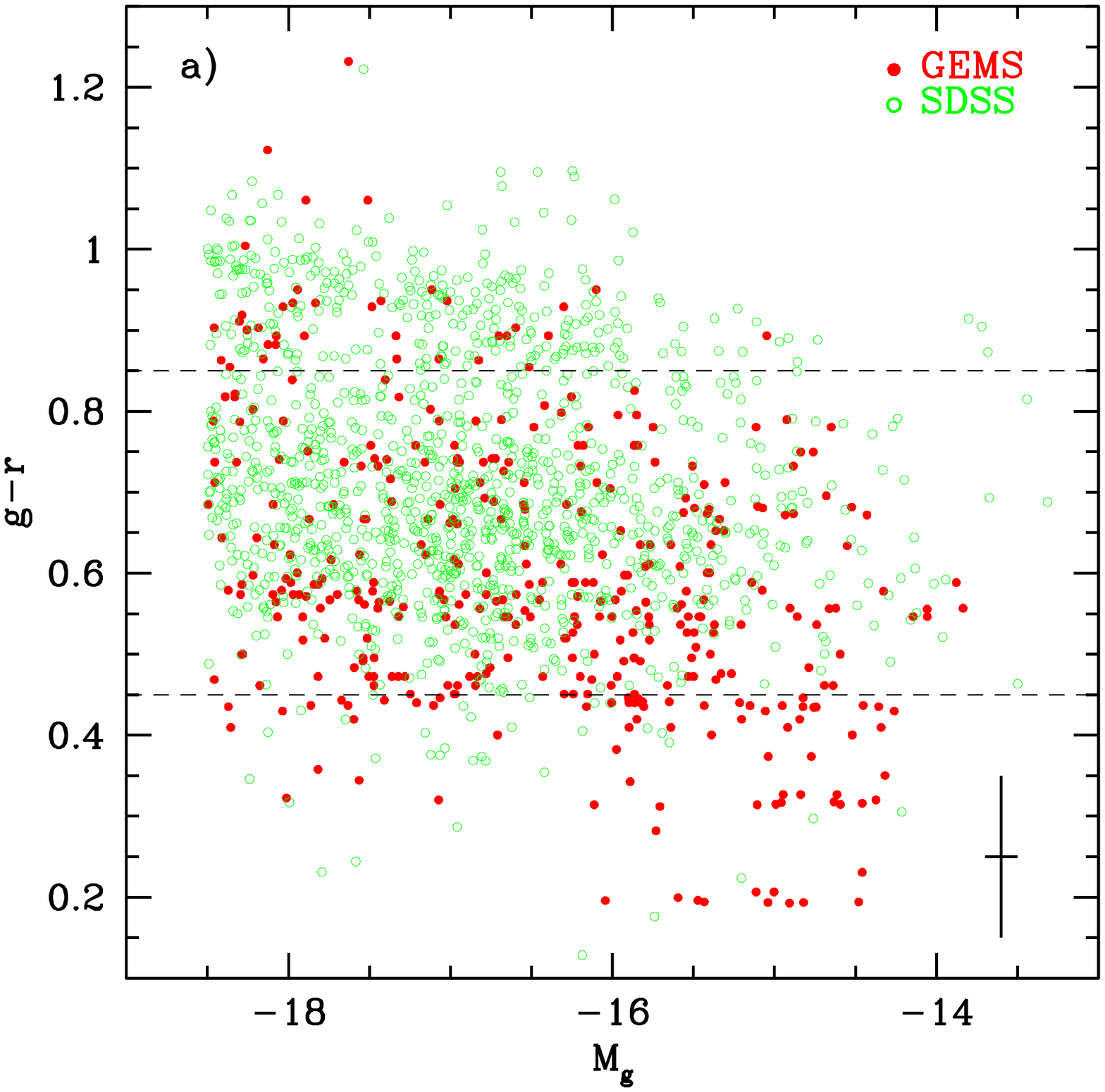}{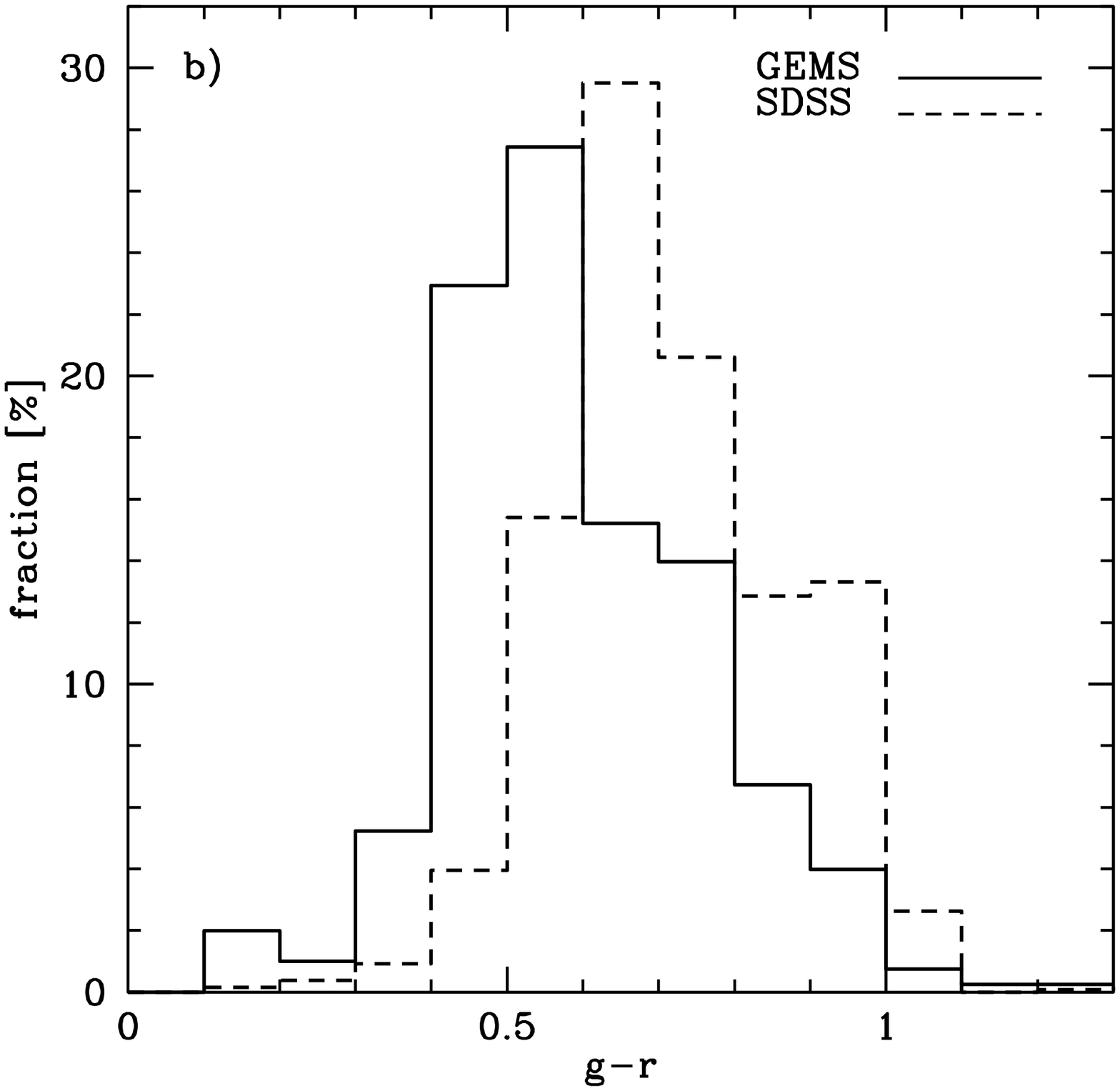}
\caption{{\bf a)} The color magnitude diagram for the GEMS and SDSS samples.
The AB magnitudes given in SDSS have been transformed to Vega magnitudes using
the transformations given in \citet{wol04}. The error bar applies to both
samples. The two dashed lines indicate the typical colors of early type dwarfs
in the Local Group \citep[upper line,][]{mat98} and of BCDs
\citep[lower line,][]{gil03}. There are specific $g-r$ values (e.g. at
$g-r=0.3$), around which the GEMS dwarfs seem to cluster, resulting in some
distinct horizontal features. These stem from the template fitting process and
are not real. The resulting gaps in the distribution are smaller than the
errors and are therefore not affecting the results. {\bf b)} The $g-r$ color
distributions for both samples. The median colors are 0.57 and 0.70 for GEMS
and SDSS, respectively.\label{col}}
\end{figure}

The Combo-17 survey provides rest-frame magnitudes in the SDSS $ugr$ bands,
which allows us to directly compare the global colors of the GEMS and SDSS
dwarf samples. The resulting color magnitude diagrams for both samples are
shown in Figure \ref{col}a. The difference in global colors between the two
samples is apparent in the histograms shown in Figure \ref{col}b.
The median colors are 0.57 and 0.70 for GEMS and SDSS, respectively.
A KS-test yields a probability of $\sim2\times10^{-41}$ that the two color
distributions stem from the same parent distribution.

In addition, there is a population of low luminosity, very blue GEMS dwarfs in
Figure \ref{col}a, with hardly any local SDSS dwarf counterparts. We refer to
these objects as low luminosity blue dwarfs (LLBDs) and include all objects
with $M_B>-16.1$ mag and $B-V<0.26$ mag to this group. This leads to 48 LLBDs
in the GEMS sample in the redshift range $z=0.09-0.23$ ($\sim12\%$ of the
sample) and 8 LLBDs in SDSS ($\sim0.6\%$ of the sample). The magnitudes and
colors of the LLBDs are consistent with a recent intermediate- to low-mass
starburst. We note, however, that the LLBDs are significantly bluer
($B-V\sim-0.1$ to $0.26$ mag) than typical blue compact dwarfs
\citep[$B-V\approx0.45$,][]{cai01}. On the other hand, there are two well
studied dwarfs with luminosities and colors similar to the LLBDs: I Zw 18
\citep{izo04,ost05} and SBS 1415+437 \citep{thu99,alo05}. However, we have to
emphasize that the number of LLBDs in GEMS might be strongly overestimated.
Some of these objects exhibit abnormally high IR emission, which indicates that
they might in fact be star forming galaxies at much higher redshifts.
Furthermore, 10 LLBDs exhibit a second smaller peak in their redshift
probability distributions in COMBO-17, which typically occurs at $z\ga1$. This
caveat has to be kept in mind, whenever we discuss LLBDs in the following
sections.

Finally, the distribution of the SDSS galaxies defines a red sequence of dwarf
galaxies, which has already been shown in the study of \citet{bla205}. These
red galaxies have been identified to be early type dwarfs, predominantly dwarf
ellipticals. It is not clear, whether the GEMS dwarfs also exhibit a red
sequence, probably because the number of objects is too small.

\subsection{Structural parameters from S\'ersic fits}\label{fitp}
\begin{figure}
\epsscale{1}
\plotone{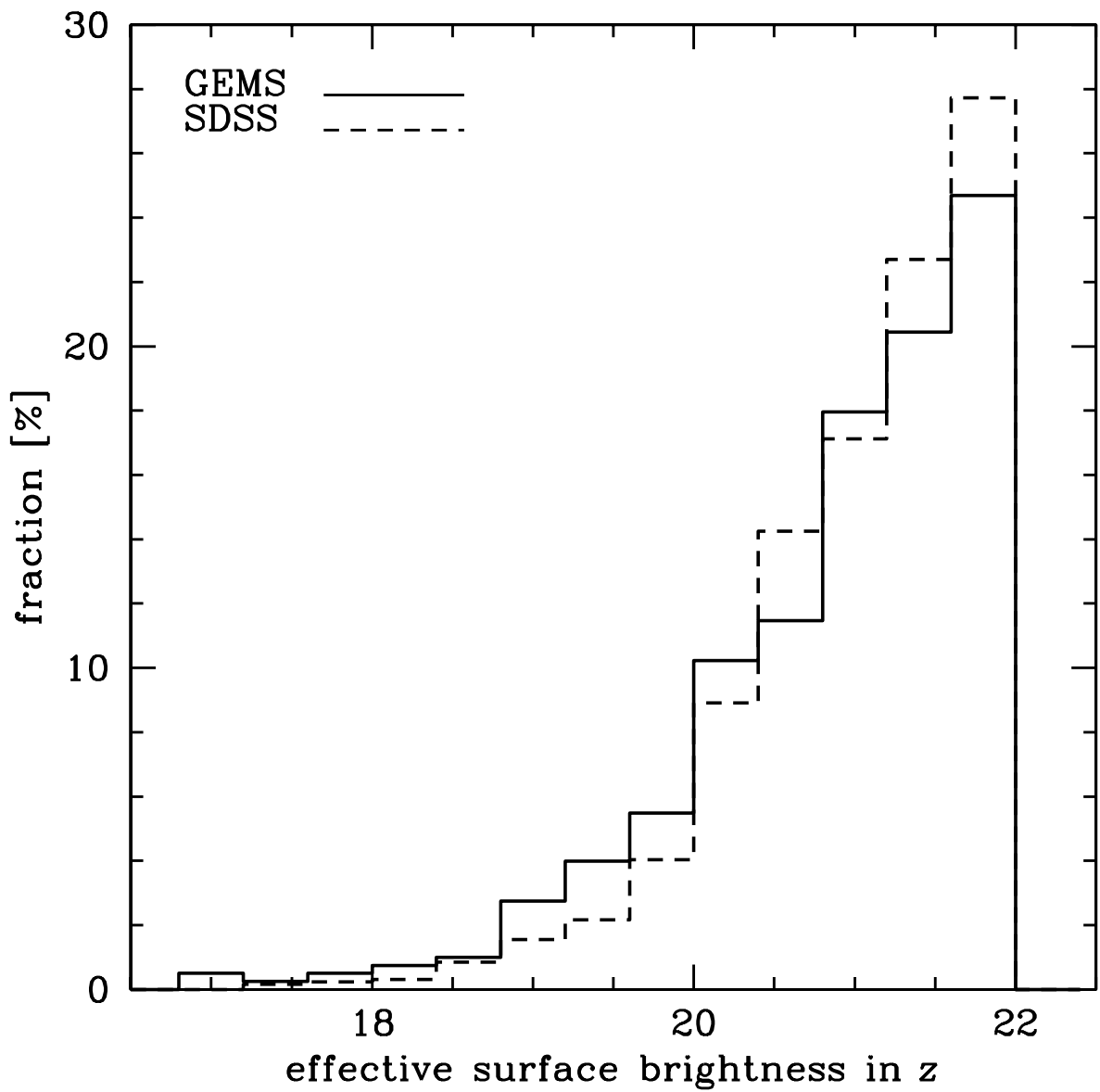}
\caption{Histogram of the effective surface brightness distribution ($\mu_e$)
for the GEMS and SDSS samples. $\mu_e$ is the mean surface brightness in $z$
within the effective (half light) radius, which was determined by a S\'ersic
fit to the $z$-band images. The median values are 21.10 mag arcsec$^{-2}$ and
21.21 mag arcsec$^{-2}$ for GEMS and SDSS, respectively.\label{esb}}
\end{figure}

We determined structural parameters such as half light radii ($r_e$) and
effective surface brightnesses ($\mu_e$) by fitting a S\'ersic model to the
light distributions of the $z$-band images for both samples. The NYU-VAGC does
provide such parameters for all objects. However, these parameters deviate
systematically from the ones for the GEMS sample, which were determined using
GALFIT. It seems that these differences have been introduced by the different
fitting procedures and weighting schemes (see Appendix). Hence, we repeated the
S\'ersic fits to the SDSS objects using GALFIT, making sure that both samples
are fitted using exactly the same procedure, software, and weighting scheme. In
Figure \ref{esb} we show the histograms of $\mu_e$ in the $z$-band for both
samples. The effective surface brightness is defined as the mean surface
brightness within $r_e$. The two distributions agree quite well and the
corresponding median values are very similar (21.10 mag arcsec$^{-2}$ for GEMS
and 21.21 mag arcsec$^{-2}$ for SDSS).

\begin{figure}
\epsscale{1}
\plottwo{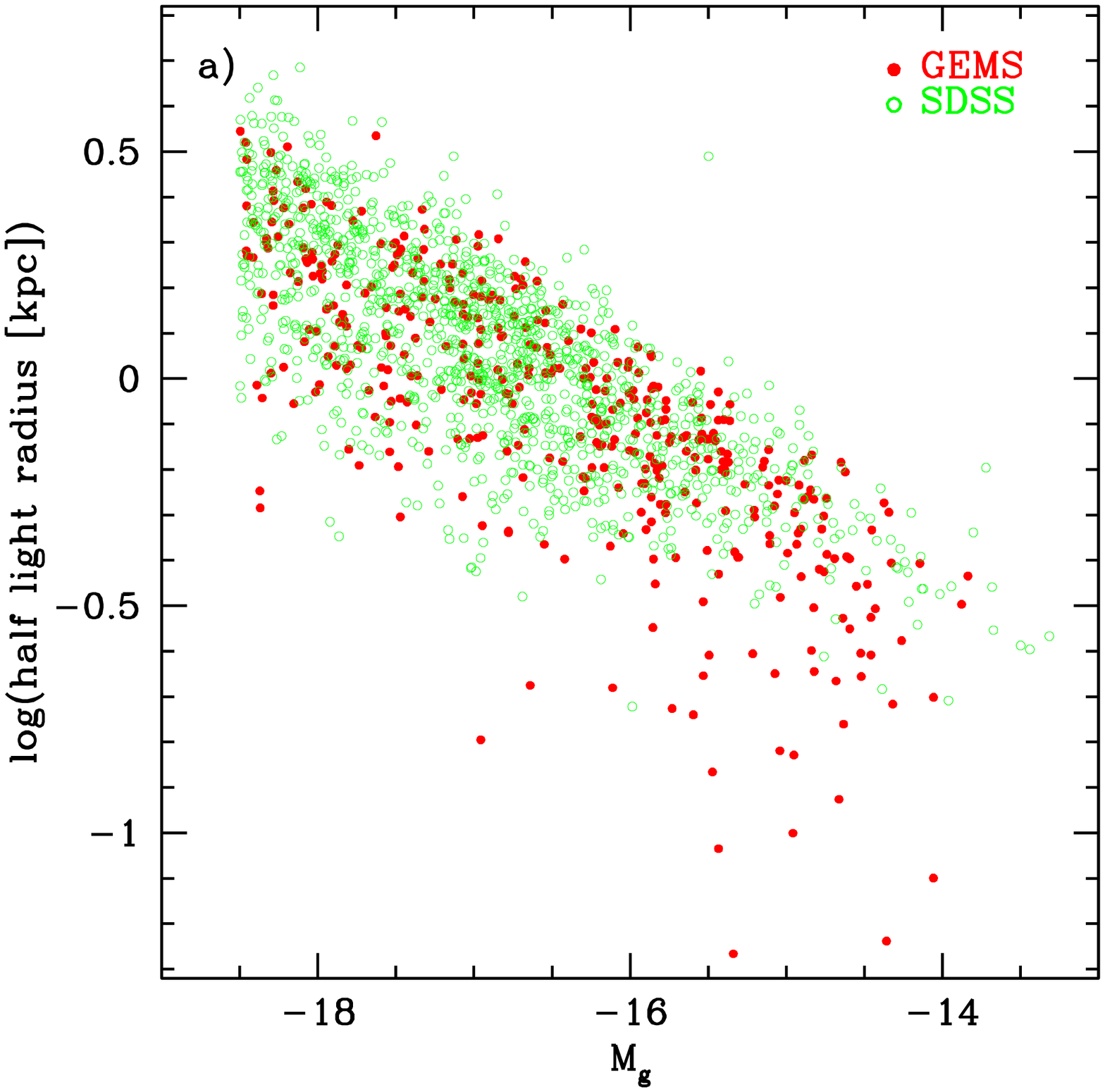}{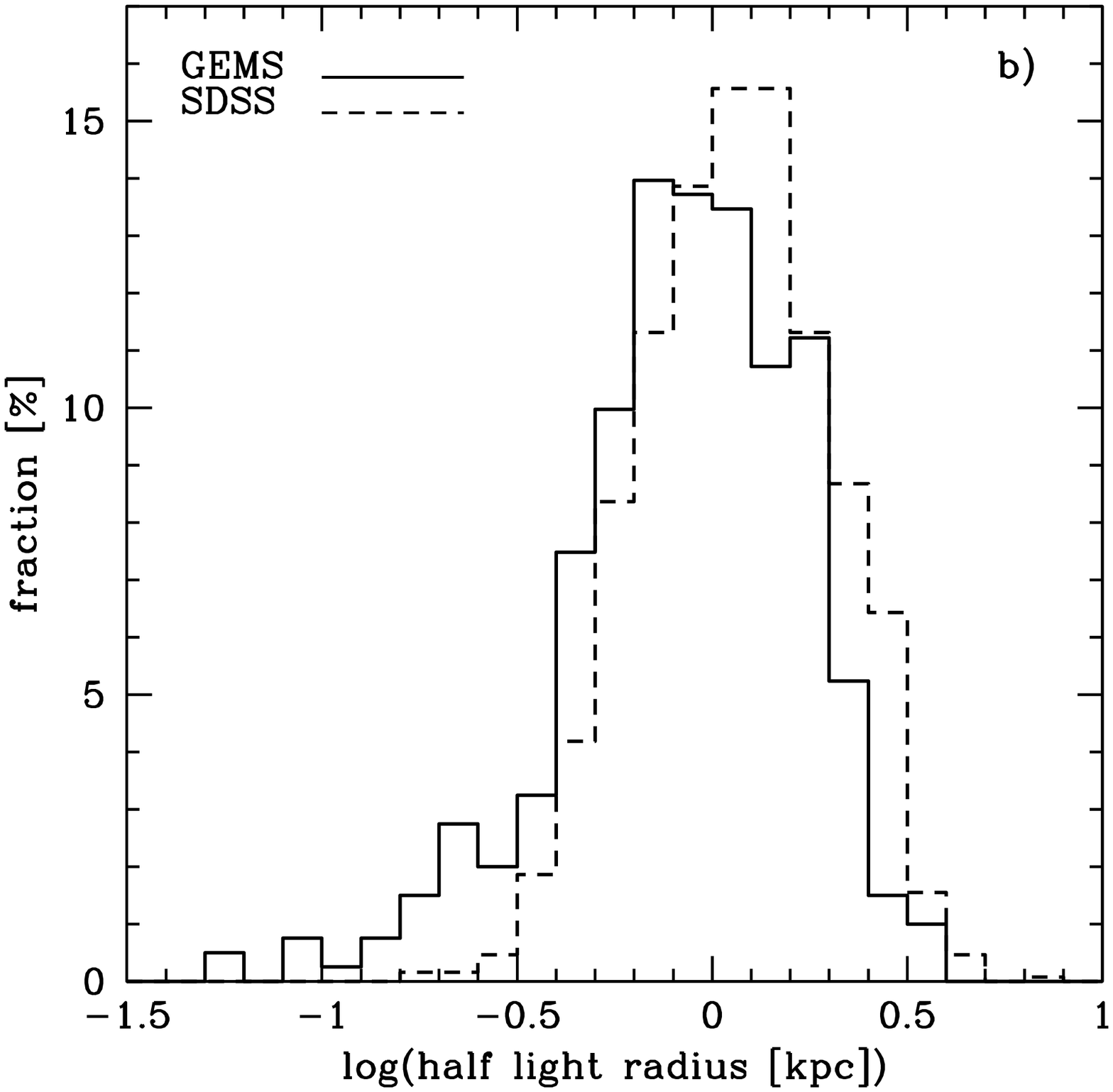}
\caption{{\bf a)} The logarithm of the half light radius in kpc versus $M_g$.
The half light radii have been determined from a S\'ersic fit to the $z$-band
images using GALFIT. A linear fit to the GEMS sample gives $-0.20\times
M_g-3.33$; the median half light radius is 892 pc. For the SDSS sample we
obtain $-0.18\times M_g-2.98$ and 1157 pc. {\bf b)} Histograms of the
logarithms of the half light radii for both samples.\label{era}}
\end{figure}

\begin{figure}
\epsscale{1}
\plotone{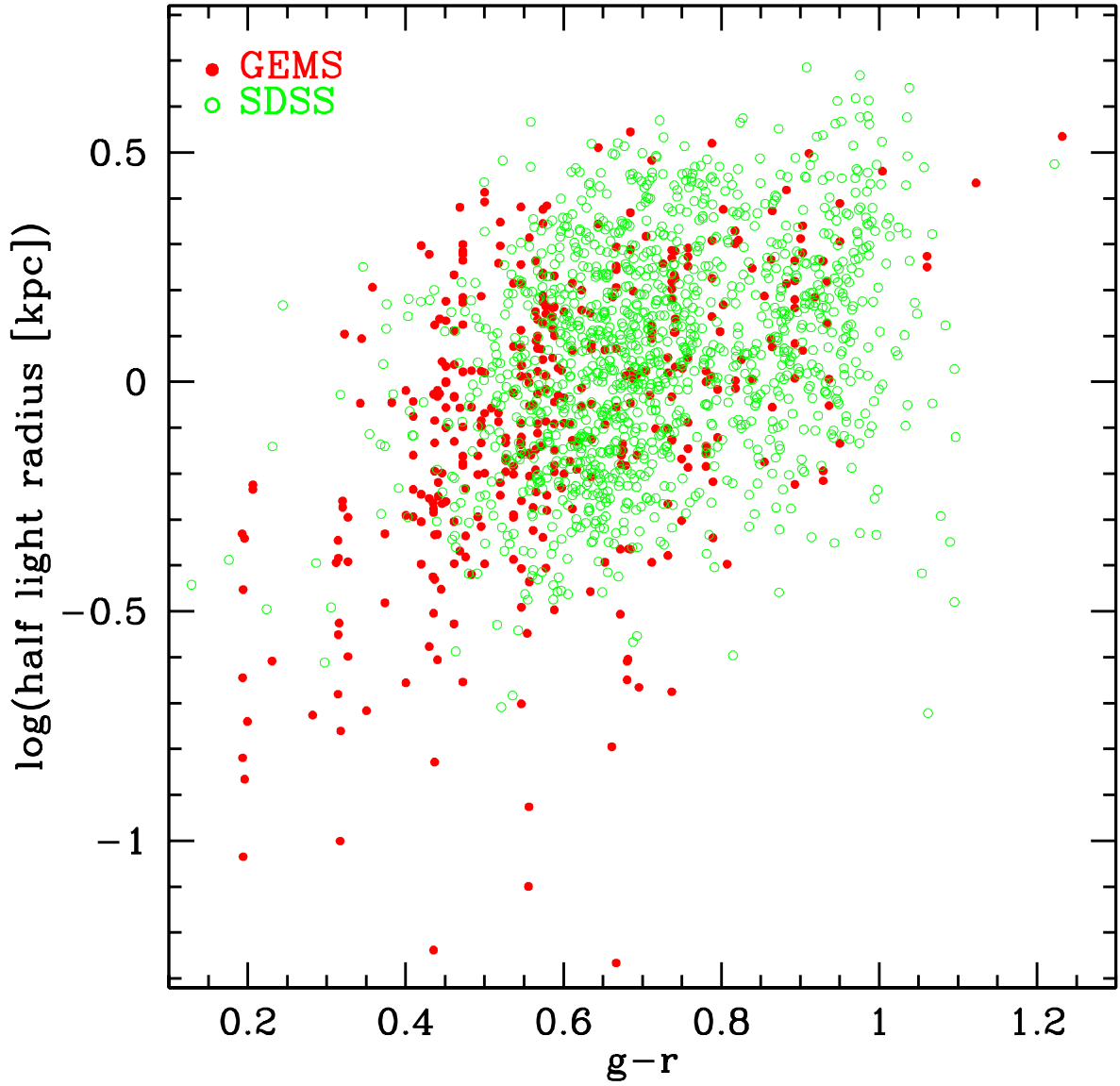}
\caption{The $g-r$ color versus the logarithm of the half light radius in
kpc.\label{era1}}
\end{figure}

In Figure \ref{era}a we plot $log(r_e)$ versus $M_g$ for both samples and
Figure \ref{era}b shows the histograms for $log(r_e)$. The distributions
overlap quite well, with the majority of GEMS dwarfs having comparable $r_e$ to
SDSS dwarfs. There is however a tail of low $r_e$ values exhibited by a small
number of GEMS dwarfs. The median $r_e$ for GEMS is 892 pc and for SDSS 1157
pc. A small contribution to this difference stems from the fact that, due to
the redshift, the $r_e$ for the GEMS dwarfs have been measured in a slightly
bluer band, which is actually closer to the $i$-band than to the $z$-band for
the median redshift. Therefore, we also fitted S\'ersic models to the SDSS
$i$-band images. The corresponding median value is 1125 pc. Thus, the
difference in median values remains and might in part be caused by the red
sequence dwarfs in SDSS (Figure \ref{col}a), which are much less prominent in
GEMS. These objects are rather bright and therefore have rather large $r_e$. On
the other hand, there is a significant population of $M_g>-16$ mag objects with
$log(r_e)$ below --0.5 in GEMS with almost no counterparts in SDSS (Figure
\ref{era}a). These are the same objects, which make up the blue low luminosity
group pointed out in $\S$ \ref{gcol} (Figure \ref{era1}).

\begin{figure}
\epsscale{1}
\plottwo{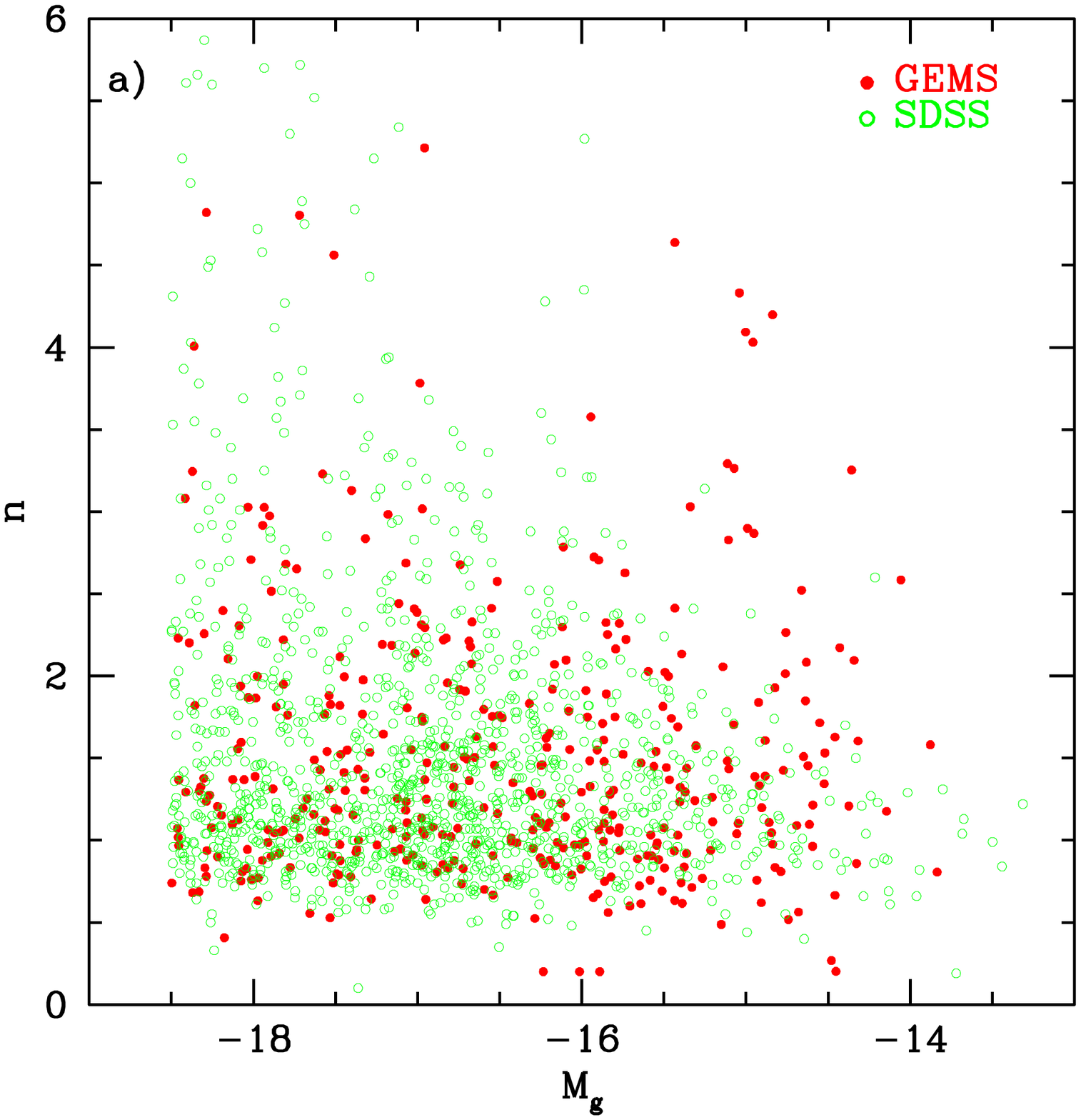}{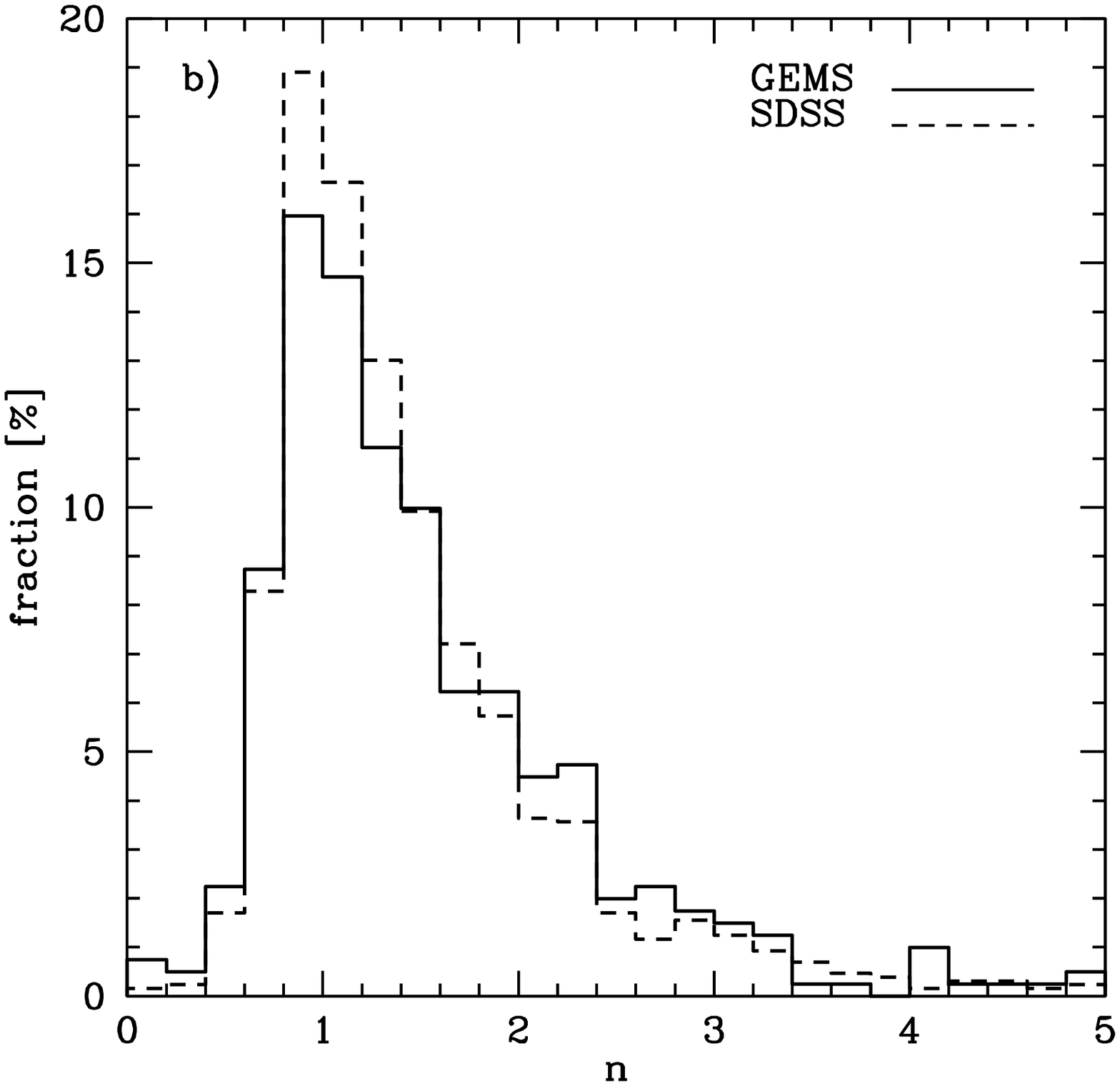}
\caption{{\bf a)} The shape parameter $n$ from the S\'ersic fits versus $M_g$.
A value of $n=1$ corresponds to a exponential disk, whereas a value of $n=4$ is
equivalent to a de Vaucouleurs model. {\bf b)} Histograms of $n$. The median
values are 1.32 and 1.25 for GEMS and SDSS, respectively.\label{sha}}
\end{figure}

Finally, in Figure \ref{sha}a the S\'ersic index $n$ is plotted versus $M_g$.
For $n=1$ the S\'ersic model is equivalent to an exponential model and for
$n=4$ it is equivalent to a de Vaucouleurs model. The two distributions are
quite similar, which is also exhibited in the histograms in Figure \ref{sha}b.
The distributions strongly peak around $n=1$ (median value for GEMS is
$n_{med}=1.32$ and for SDSS $n_{med}=1.25$), which indicates that the surface
brightness profiles of most of the dwarfs are close to an exponential model. In
fact, an exponential model is commonly found to best characterize the surface
brightness profiles of late-type dwarfs \citep{mak99,bar01,par02}. On the other
hand, for early-type dwarfs models with higher S\'ersic indices seem to be more
appropriate \citep{gra03,bar03}. This provides some evidence that the GEMS and
the SDSS samples are mainly composed of late-type dwarfs. There is also a group
of low luminosity dwarfs in GEMS with quite high $n$. These are mainly the
LLBDs, already identified in the previous sections. Hence, while the majority
of GEMS and SDSS dwarfs have S\'ersic index $n<2$, the LLBDs are more compact
and have therefore higher $n$.

\subsection{Star formation histories of GEMS dwarfs}\label{cod}
\begin{figure}
\epsscale{1}
\plotone{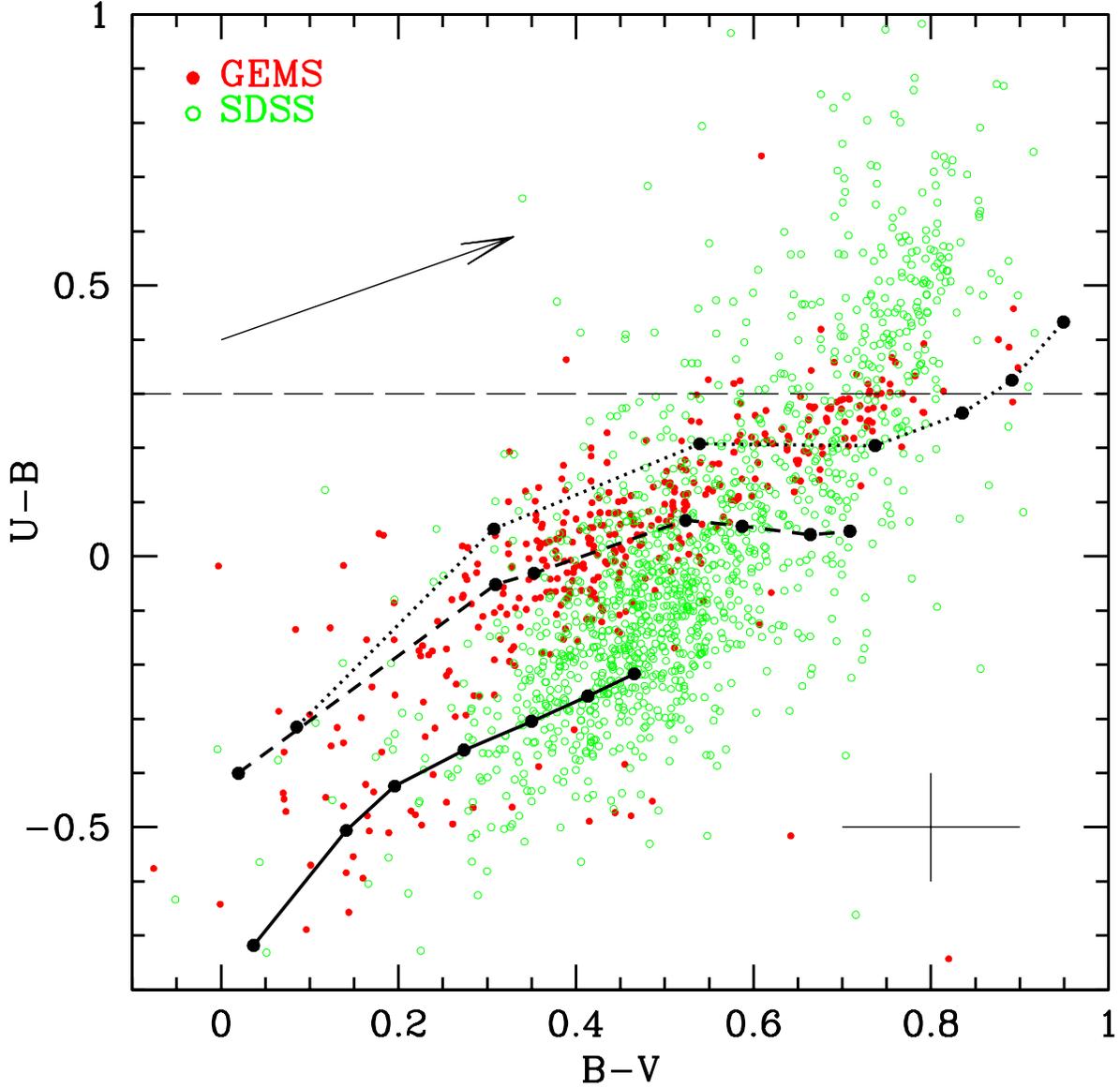}
\caption{A $B-V$ versus $U-B$ color-color plot. Overplotted are three {\it
Starburst 99} model tracks, which are all computed using a Kroupa IMF. The
black dots indicate the following time steps (from left to right): 0.1, 0.5,
1.0, 2.0, 4.0, 8.0, 15.0 Gyr. {\it Solid line:} constant star formation with
$SFR=0.03$ M$_{\sun}$ yr$^{-1}$ and $Z=0.004$. {\it Dashed line:} single
starburst with a mass of $3\times10^8$ M$_{\sun}$ and $Z=0.0004$.
{\it Dotted line:} single starburst with a mass of $3\times10^8$ M$_{\sun}$ and
$Z=0.008$. Objects above the dashed line have uncertain $\ub$ colors and are
excluded from the discussion. The error bar represents only the error from the
measurement and does not include uncertainties stemming from the color
transformation for SDSS. The arrow indicates the effect of dust on the colors
\citep{sch98}.\label{ccp1}}
\end{figure}

\begin{figure}
\epsscale{1}
\plotone{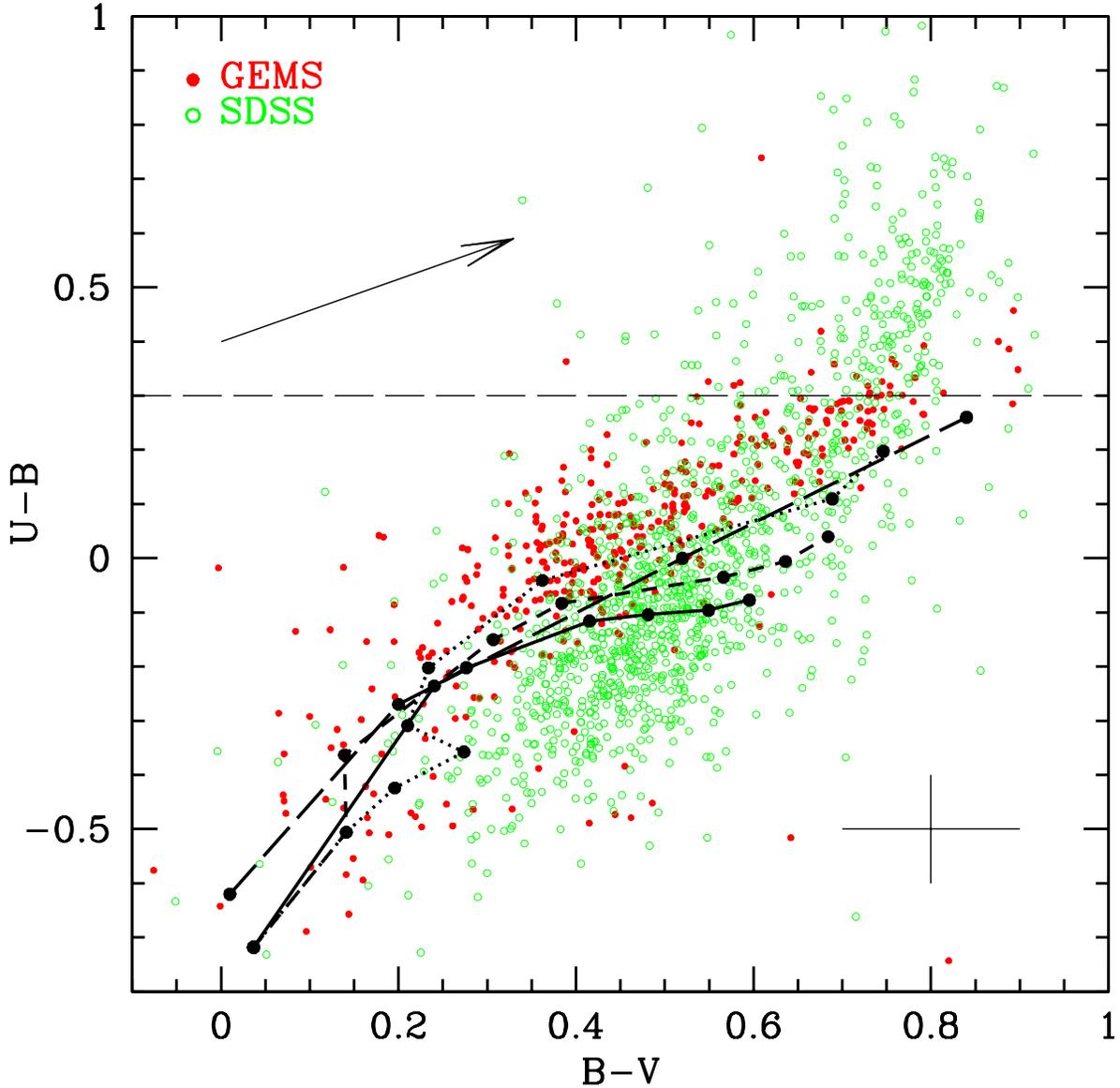}
\caption{The same as Figure \ref{ccp1}. Three lines represent models, where a
continuous star formation with a constant rate of $SFR=0.03$ M$_{\sun}$
yr$^{-1}$ and a metallicity of $Z=0.004$ has been combined with various single
starbursts starting at different times. For these models a Kroupa IMF has been
used. The fourth line represents a model with an exponentially decreasing star
formation rate and a Salpeter IMF. {\it Solid line:} A single starburst with a
mass of $3\times10^8$ M$_{\sun}$ and $Z=0.0004$ starts at 0.1 Gyr. The time
steps are the same as in Figure \ref{ccp1}. {\it Short dashed line:} A single
starburst with a mass of $3\times10^8$ M$_{\sun}$ and $Z=0.004$ starts at 0.9
Gyr. The time steps are: 0.1, 0.5, 1.0, 1.1, 1.5, 2.0, 4.0, 8.0, 15.0.
{\it Dotted line:} A single starburst with a mass of $5\times10^8$ M$_{\sun}$
and $Z=0.02$ starts at 3.9 Gyr. The time steps are: 0.1, 0.5, 1.0, 2.0, 4.0,
4.1, 4.5, 8.0, 15.0. {\it Long dashed line:} An exponentially decreasing SFR,
using models from  \citet{bru93}. The time steps are: 0.1, 1.0, 3.0, 10.0 Gyr.
The arrow indicates the effect of dust on the colors
\citep{sch98}.\label{ccp2}}
\end{figure}

In order to study the colors and star formation histories of the GEMS and SDSS
dwarf samples, we compare their $\ub$ and $\bv$ colors to model tracks
constructed using {\it Starburst 99} \citep{lei99,vaz05}. We transform the SDSS
colors to the Johnson colors using the following equations valid for the galaxy
type Im: $(\ub)=(u-g)-0.99$ and $(\bv)=(g-r)+0.07$ \citep{fuk95}. Since we did
not derive morphological classes for our samples, we cannot transform the
colors according to galaxy types. This introduces some uncertainty to the
Johnson colors of the SDSS galaxies. However, as shown by the distribution of
S\'ersic indices ($\S$ \ref{fitp}), both samples are strongly dominated by
late-type dwarfs (galaxy types Sm, Im). In addition, we are able to roughly
identify early type dwarfs according to their colors and have therefore some
control of the uncertainties. Assuming that most galaxies with $g-r\ga0.85$ are
early-types, the color transformation discussed above adds an error to their
$\ub$ color as large as 0.36. In addition, there is another error source
introduced by a red leak to the $u$ filter and a bias in the sky subtraction
reported on the SDSS web pages.\footnote{
http://www.sdss.org/dr2/start/aboutdr2.html\#imcaveat} Both uncertainties will
predominantly affect galaxies with $\bv\ga0.6$ and, in particular, affect their
$\ub$ color, causing it to be too red. Therefor, the star formation models are
not expected to be able to reproduce the corresponding $\ub$, $\bv$ color
combinations. Thus, we will exclude all SDSS galaxies with $\ub>0.3$ (see
Figure \ref{ccp1}) from the following discussion (242 objects, $\sim19\%$ of
the SDSS sample), since their $\ub$ colors are too uncertain. The line
$\ub=0.3$ is drawn on Figure \ref{ccp1} and \ref{ccp2} for reference.

In Figure \ref{ccp1} the $\ub$ versus $\bv$ color-color plot is shown. A
general difference between the two samples is again obvious. In order to
understand, whether this difference could be linked to the evolution of the
galaxies over $\sim2$ Gyr (corresponding to the look-back time of the GEMS
sample), we compare the colors to different {\it Starburst 99} model tracks.
All models shown are based on a Kroupa Initial Mass Function
\citep[IMF,][]{kro01}. The solid line in Figure \ref{ccp1} represents a model
with a constant star formation rate (SFR) of $0.03$ M$_{\sun}$ yr$^{-1}$ and a
metallicity of $Z=0.004$. These are typical values for field late type dwarfs
\citep{zee01,hun04}. The dashed line shows the color evolution of a Single
Stellar Population (SSP) with a mass of $3\times10^8$ M$_{\sun}$ and a
metallicity of $Z=0.0004$ formed in a single starburst. The dotted line
represents the same SSP, but with a metallicity of $Z=0.008$. The solid dots
mark the following time steps (from left to right): 0.1, 0.5, 1.0, 2.0, 4.0,
8.0, 15.0 Gyr.

All three models seem to be able to account for the colors of specific
subgroups in Figure \ref{ccp1}. However, they are most likely too simple, in
order to represent the star formation histories of these dwarfs. The model with
constant star formation does reproduce the colors of the bluest objects in both
samples, but its luminosity remains too low ($M_B\approx-12$ mag) and we would
have to apply an unreasonably high SFR ($>0.5M_{\sun}$ yr$^{-1}$) to reach the
observed luminosities. The single burst models are able to cover the colors of
the redder objects in Figure \ref{ccp1} and they should be representative of
galaxies, which do not form stars at the present time, i.e. of early type
dwarfs. However, the stellar populations of early type dwarfs in the Local
Group, the only such objects which can be studied in enough detail, are known
to be more complex than a SSP \citep{gre98,iku02,tol03}. It is therefore
reasonable to assume that the star formation histories of dwarfs constitute a
mixture of multiple bursts and periods of constant star formation, probably
with different SFRs. Figure \ref{ccp1} suggests that the combination of
different modes of star formation can reproduce the observed colors.

In Figure \ref{ccp2} we plot tracks of models, which combine a single starburst
with continuous star formation. The onset of the burst occurs at different
times: together with the star formation at the earliest time step (solid line),
0.9 Gyr after the beginning of the continuous star formation (short dashed
line), and 3.9 Gyr after the beginning of the continuous star formation (dotted
line). In addition, we add a model with an exponentially decreasing SFR (long
dashed line). These models are still just an approximation to the real star
formation histories. For instance, the starburst component of the star
formation histories could also be represented by several minor bursts, instead
of one larger burst. However, the models reproduce the required luminosity
range ($-18\la M_B\la-14$ mag between roughly 4--15 Gyr) and they cover the
observed colors. The most important aspect of these model tracks is that they
all exhibit a certain period of time, in which the $\ub$ color remains more or
less constant, but the $\bv$ color changes significantly. This is in
qualitative agreement with the $\bv$ color difference between the samples shown
in Figures \ref{ccp1} and \ref{ccp2}, and also with the different distributions
in Figure \ref{col}b.

The models also show that an age difference of $\sim2$ Gyr is enough to
increase the $\bv$ color by $\sim0.1$, which is of the same order as observed.
In addition, the color changes more rapidly at earlier times than at later
times. This property is mostly caused by the burst component, whereas the
passive evolution after the burst changes the colors only weakly.
Interestingly, this characteristic is also shown by the observations. The
median $\bv$ color difference between the samples decreases with increasing
$\ub$ color. For the color bin $-0.5<\ub<-0.3$ the difference is 0.16 and then
decreases along the distribution reaching 0.01 for the bin $0.1<\ub<0.3$. This
indicates that the sequence of galaxies in the color-color plot is mainly due
to the age of the most recent episode of star formation. The bluest galaxies
are those having experienced a recent episode of star formation, while the
redder dwarfs are older remnants. The luminosity appears to be less important
for this relation, which is also indicated by the shallow slope in the color
magnitude diagram (Figure \ref{col}a) and its large scatter.

We can therefore conclude that the range of global colors shown by GEMS and
SDSS dwarfs (over look-back times of 3 Gyr) is consistent with a star formation
history involving multiple bursts of star formation, possibly combined with
intermediate periods of relatively constant SFRs. Also a model representing an
exponentially decreasing SFR is able to reproduce the observed colors. In
particular, we note that a single burst model or a history of constant star
formation alone, cannot reproduce the full color range. We also find that the
bluer colors typical of GEMS dwarfs (present at $z=0.09-0.25$ or look-back
times of 1--3 Gyr) can evolve by $\sim0.1$ mag into the redder colors of SDSS
dwarfs (present at $z<0.02$) over $\sim2$ Gyr.

\subsection{Feedback from star formation and requirements for blowaway in
dwarfs}
\begin{figure}
\epsscale{1}
\plotone{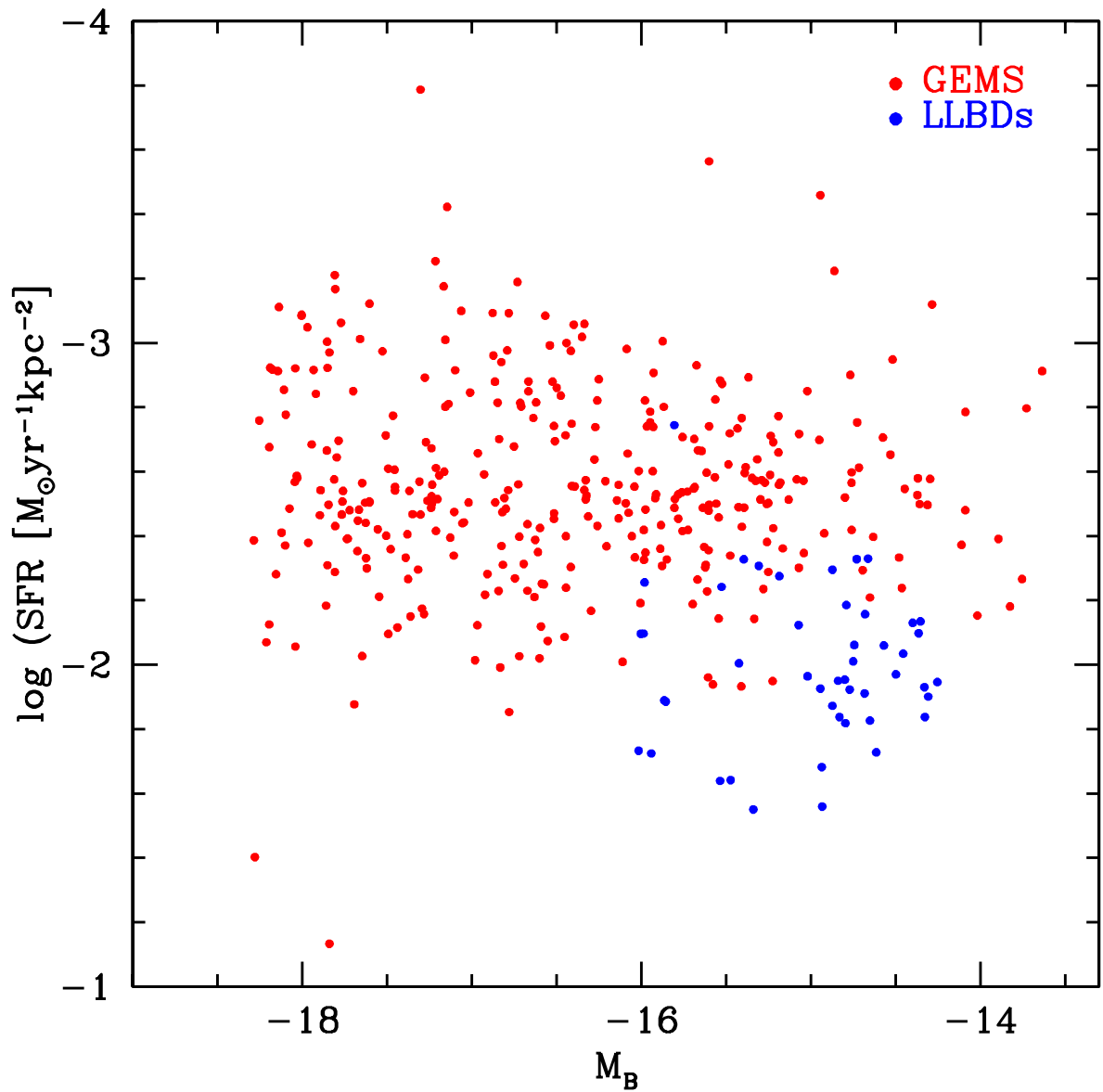}
\caption{The normalized SFR versus $M_B$. The SFRs have been estimated from the
2800\AA~continuum fluxes ($L_{2800}$) and using the equation
$SFR$[M$_{\sun}$yr$^{-1}]=3.66\times10^{-40}L_{2800}$ [ergs
s$^{-1}~\lambda^{-1}]$ adopted from \citet{ken98}. These SFRs have then been
divided by the isophotal area provided by Sextractor.\label{sfr}}
\end{figure}

Empirical constraints on the luminosity, star formation histories, and star
formation feedback in dwarf galaxies at different look-back times can provide
useful constraints for hierarchical Lambda-CDM models of galaxy formation and,
in particular, on the issue of feedback and blowout of gas via star formation
in low mass galaxies. In a first step we, therefore, estimate the SFRs of the
GEMS dwarfs. The COMBO-17 survey provides the rest-frame luminosities in a
synthetic UV band centered on the 2800\AA~line. For the redshift range of our
GEMS sample the luminosities in this band are based on extrapolations beyond
the filter set used in COMBO-17. For the luminosities in this UV band, we can
derive the fluxes in the 2800\AA~line, $L_{2800}$. In order to estimate the
SFRs we then use the following equation \citep{ken98}:
\[SFR~[M_{\sun}yr^{-1}]=3.66\times10^{-40}L_{2800}~[ergs~s^{-1}~\lambda^{-1}]\]
This equation applies to galaxies with continuous star formation over the last
$10^8$ years, which is likely the case for a majority of our dwarfs. For
objects, which experienced a starburst within the last $10^8$ years, the SFRs
obtained in this way will underestimate the actual SFR. In Figure \ref{sfr} we
plot the {\it normalized} SFR versus $M_B$. For the normalization we use the
isophotal area provided by Sextractor. The range of normalized SFRs we obtained
by this rough estimate agrees very well with the values found for a sample of
nearby field dwarfs presented by \citet{hun04}. Figure \ref{sfr} indicates that
the LLBDs have relatively high SFRs per unit area.

In order to obtain an estimate of the energies needed for an ejection of gas
via starbursts in dwarf galaxies we consider the model of \citet{mac99}. The
study estimates the impact of repeated supernovae on the interstellar medium.
Two specific scenarios are distinguished: blowout, in which gas is expelled
from the disk, but will in part be reaccreted at later epochs, and blowaway, in
which the gas becomes completely unbound and is lost to the dwarf. The latter
scenario is the one we consider for our estimate. The mechanical luminosity
needed for a blowaway to occur depends on the visible mass of the dwarf and the
effective sound speed, which in turn is related to the axis ratio
\citep{mac99}. The corresponding equations are:
\[L_{38}>8\times10^{-2}M_{vis,7}^{7-6\alpha}\left(\frac{\phi}{\omega_0} 
\right)^6c_{10}^{-10}h\]
and
\[\epsilon=0.43\frac{\phi}{\omega_0}M_{vis,7}^{1-\alpha}c_{10}^{-2}\]

\begin{figure}
\epsscale{1}
\plotone{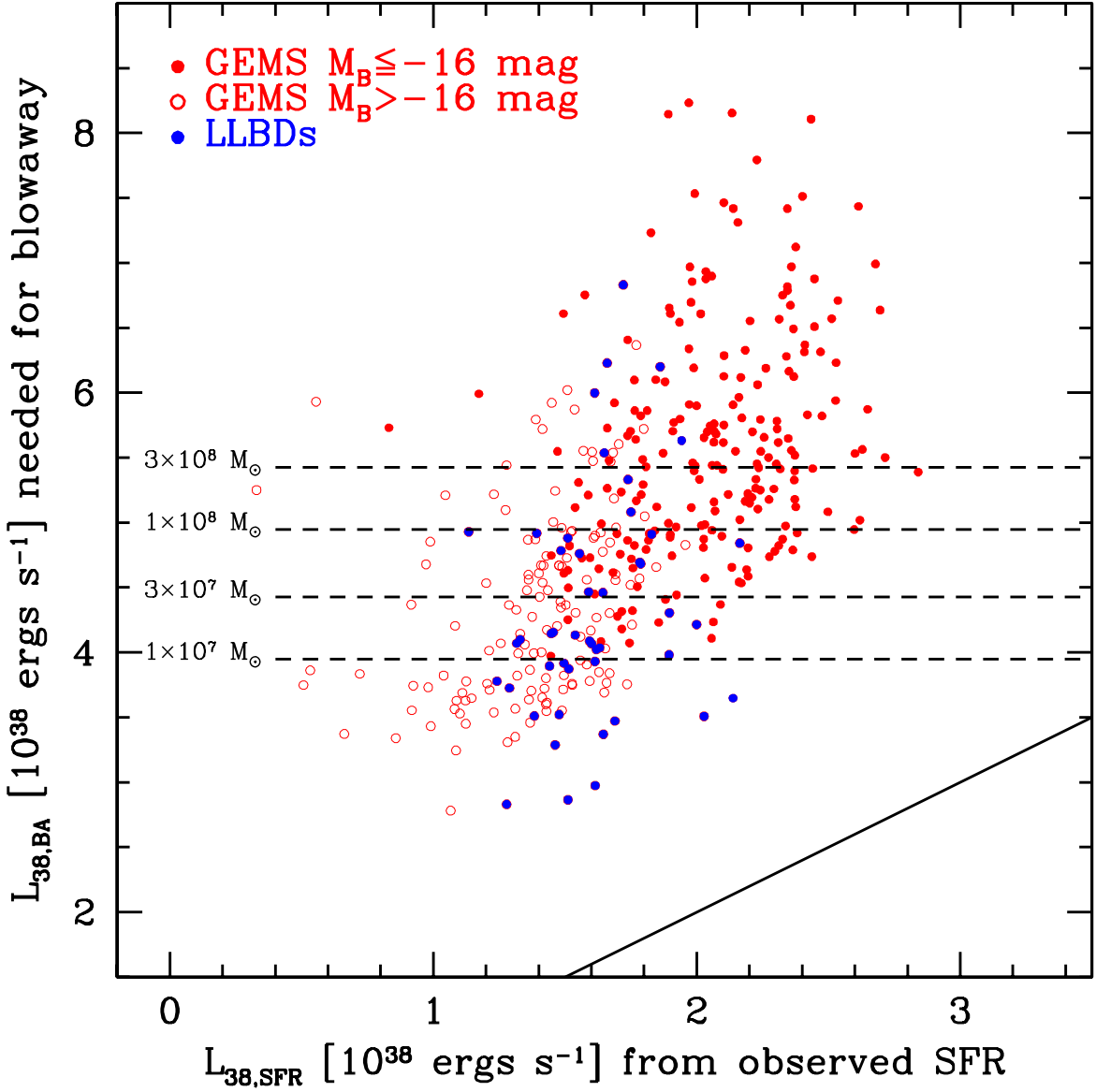}
\caption{Plot of the mechanical luminosity needed for a complete blowaway of
the gas in dwarfs having a given $M_V$ and a range of apparent ellipticities
versus the mechanical luminosities inferred from the SFRs. $L_{38}$ corresponds
to the logarithm of the mechanical luminosity in units of $10^{38}$ergs
s$^{-1}$. The four dashed lines mark the peak mechanical luminosities reached
of starbursts with the indicated masses. The solid line corresponds to
$L_{38,SFR}=L_{38,BA}$.\label{l38}}
\end{figure}

$L_{38}$ corresponds to the logarithm of the mechanical luminosity in units of
$10^{38}$ergs s$^{-1}$, $M_{vis,7}$ is the visible mass in units of
$10^7M_{\sun}$, $\alpha=0.338$ is a constant, $\phi$ is the dark-to-visible
mass ratio given as $\phi\backsimeq34.7M_{vis,7}^{-0.29}$, $\omega_0=3$ is a
constant, $c_{10}$ is the effective sound speed in units of 10 km s$^{-1}$,
$h=H_0/100$, where $H_0$ is the Hubble constant, and $\epsilon$ is the
ellipticity. We estimate the visible mass of our galaxies using the $V$-band
luminosity. The axis ratios are provided by the S\'ersic fits. It is clear that
more luminous galaxies need a higher mechanical luminosity for complete
blowaway of the gas. This is due to the fact that we used the luminosity to
derive the mass, which was then used to determine the mechanical luminosity.
For a given $M_V$, there is a range in mechanical luminosities for blowaway due
to the range in ellipticities. Note, however, that the mechanical luminosities
obtained in this way have to be considered as lower limits, since we most
likely underestimate the mass and use apparent ellipticities. The estimates of
the mechanical luminosities needed for a complete blowaway of the gas in our
dwarfs ($L_{38,BA}$) is plotted versus the mechanical luminosities inferred
from the SFRs ($L_{38,SFR}$) in Figure \ref{l38}. The solid line corresponds to
$L_{38,SFR}=L_{38,BA}$. Obviously, no significant gas loss would occur in these
objects, if their SFRs would never exceed the ones inferred from their observed
near-UV luminosities.

Next, using the star formation histories and starburst strengths that were
found to reproduce the observed range in colors of GEMS and SDSS dwarf galaxies
(see $\S$ \ref{cod} and Figure \ref{ccp2}), we computed the maximum mechanical
luminosities, which can be injected into the ISM by the relevant starbursts of
different masses. The range in luminosities is illustrated on Figure \ref{l38}
by three dashed lines corresponding to starburst masses of $3\times10^8$
M$_{\sun}$, $1\times10^8$ M$_{\sun}$, $3\times10^7$ M$_{\sun}$, and
$1\times10^7$ M$_{\sun}$. We find that for their derived star formation
histories, the luminous ($M_B=-18$ to $-16$ mag) dwarfs are likely to retain
their gas and avoid blowaway. However, there are a fair number of low
luminosity dwarfs ($M_B=-14$ to $-16$) that are susceptible to a complete
blowaway of gas, {\it if they were} to experience a starburst. However, in
practice, only a small fraction of these low luminosity dwarfs {\it may be
actually undergoing} a starburst. Even though, we do not have any clear
evidence that some dwarfs in our sample experience a starburst at the time of
observation, we are also not able to rule this out. The derived mechanical
luminosities stem from the SFRs, which have been determined assuming that the
dwarfs had a constant SFR over the last $10^8$ years. In addition, we used the
near-UV luminosities, which could be affected by dust. In view of these
uncertainties, the derived SFRs have to be considered as lower limits. The
LLBDs could actually be a population, for which the SFRs have been
significantly underestimated, since their magnitudes and colors are consistent
with a recent low to intermediate starburst ($\S$ \ref{gcol}) and they are
compact (Figure \ref{era1}), such that they will likely have a high SFR per
unit area. This is confirmed in Figure \ref{sfr}.

\section{Summary and Conclusions}\label{sum}
We present a study of the colors, structural properties, and star formation
histories in a sample of $\sim1600$ dwarfs present over the last 3 Gyr
($z=0.002-0.25$). Our sample consists of 401 dwarfs over $z=0.09-0.25$
(corresponding to $T_{back}$ of 1 to 3 Gyr) from the GEMS survey and a
comparison sample of 1291 dwarfs with $z<0.02$ ($T_{back}<0.3$ Gyr) from the
SDSS. Our final sample is complete down to an effective surface brightness of
22 mag arcsec$^{-2}$ in $z$ and includes dwarfs with $M_g=-18.5$ to $-14$ mag.
The main results are:

The mean global color of GEMS dwarfs ($g-r=0.57$ mag) is bluer than that of the
SDSS dwarfs ($g-r=0.70$ mag) by 0.13 mag. Using {\it Starburst 99}, we find
that the full range of global colors shown by SDSS and GEMS dwarfs, over
look-back times of 3 Gyr, is consistent with a star formation history
involving bursts of star formation, combined with periods of relatively
constant SFRs or exponentially decreasing SFRs. In particular, we note that a
single burst model or a history of constant star formation alone, cannot
reproduce the full color range. We also find that the bluer colors typical of
GEMS dwarfs can evolve by $\sim0.1$ mag into the redder colors of SDSS dwarfs
over $\sim2$ Gyr.

We identify a population of low luminosity ($M_B=-16.1$ to $-14$ mag), blue
($\bv<0.26$ mag) dwarfs (LLBDs) in the GEMS sample, with hardly any
counterparts among local SDSS dwarfs. The very blue colors of the LLBDs are
comparable to those of systems, such as I Zw 18 and SBS 1415+437. Their
magnitudes and colors are consistent with a recent intermediate-to-low mass
starburst. However, we have to stress the caveat that a large fraction of these
objects might in fact be star forming galaxies at much higher redshifts. This
is indicated by an abnormally high IR emission of some of these objects and the
fact, that 10 LLBDs (out of 48) exhibit a second peak around $z\sim1$ in their
redshift probability distributions.

We performed S\'ersic fits to the GEMS and SDSS dwarfs using GALFIT making sure
that both samples are fitted using the exact same procedure and weighting
schemes. Our experience shows that this approach is essential for avoiding
large spurious differences (see Appendix). We find that $\sim76\%$ of GEMS
dwarfs and $\sim81\%$ of SDSS dwarfs have S\'ersic $n<2$. The majority of GEMS
dwarfs have half light radii comparable to those of SDSS dwarfs.

We estimate the SFR per unit area of GEMS dwarfs using the rest-frame UV
(2800\AA) luminosity and find values in the range $5\times10^{-3}$ to
$5\times10^{-1}$ M$_{\sun}$ yr$^{-1}$ kpc$^{-2}$. We note that the LLBDs have
magnitudes and colors consistent with a recent low-to-intermediate mass
starburst, small half light radii, and resultant high SFRs per unit area.

Finally, we estimate the mechanical luminosities needed for the gas in the GEMS
dwarfs to become unbound and lost (blowaway). We then compare these to maximum
mechanical power that would be injected by starbursts that are consistent with
the derived star formation histories of these dwarfs. We find that the luminous
($M_B=-18$ to $-16$ mag) dwarfs are likely to retain their gas and avoid
blowaway. However, there are a fair number of low luminosity dwarfs ($M_B=-16$
to $-14$ mag) that are susceptible to a complete blowaway of gas, {\it if they
were} to experience a starburst.

\acknowledgments
F.D.B. and S.J. acknowledge support from the National Aeronautics and Space
Administration (NASA) LTSA grant NAG5-13063 and from HST-GO-10395 and
HST-GO-10428. E.F.B. was supported by the European Community's Human Potential
Program under contract HPRN-CT-2002-00316 (SISCO). C.W. was supported by a
PPARC Advanced Fellowship. D.H.M acknowledges support from the NASA LTSA Grant
NAG5-13102. C.Y.P is grateful for support by the Institute Fellowship at STScI.
Support for GEMS was provided by NASA through number GO-9500 from the Space
Telescope Science Institute STScI, which is operated by the Association of
Universities for Research in Astronomy, Inc. AURA, Inc., for NASA, under
NAS5-26555.

\appendix
\section{Systematic differences in S\'ersic parameters derived using different
fitting methods}
\begin{figure}
\epsscale{1}
\plotone{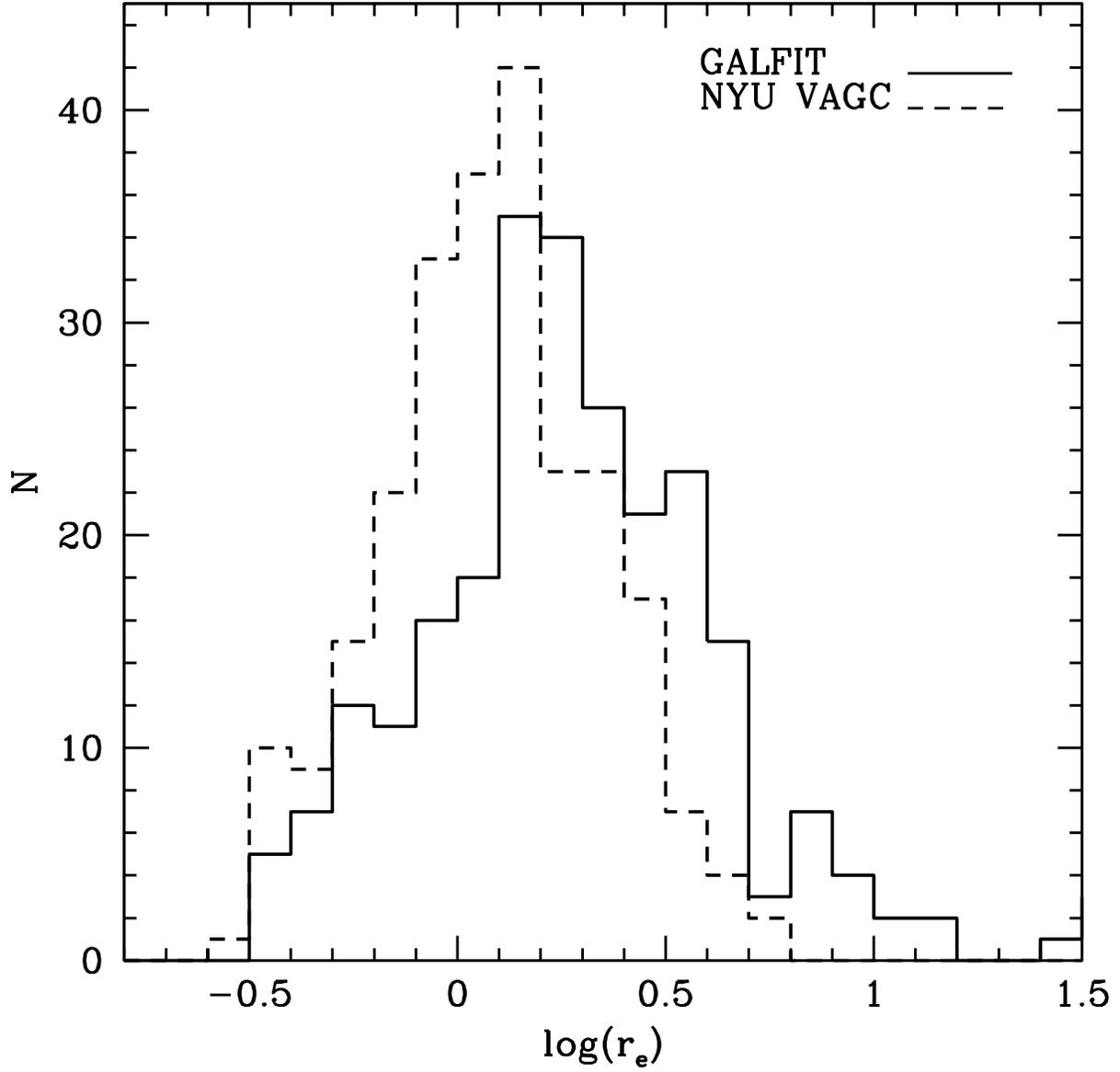}
\caption{Histograms of $r_e$ for a subsample of 245 galaxies from SDSS. The
values obtained by a S\'ersic fit using GALFIT are compared to the values
provided by the NYU VAGC.\label{ap1}}
\end{figure}

\begin{figure}
\epsscale{1}
\plotone{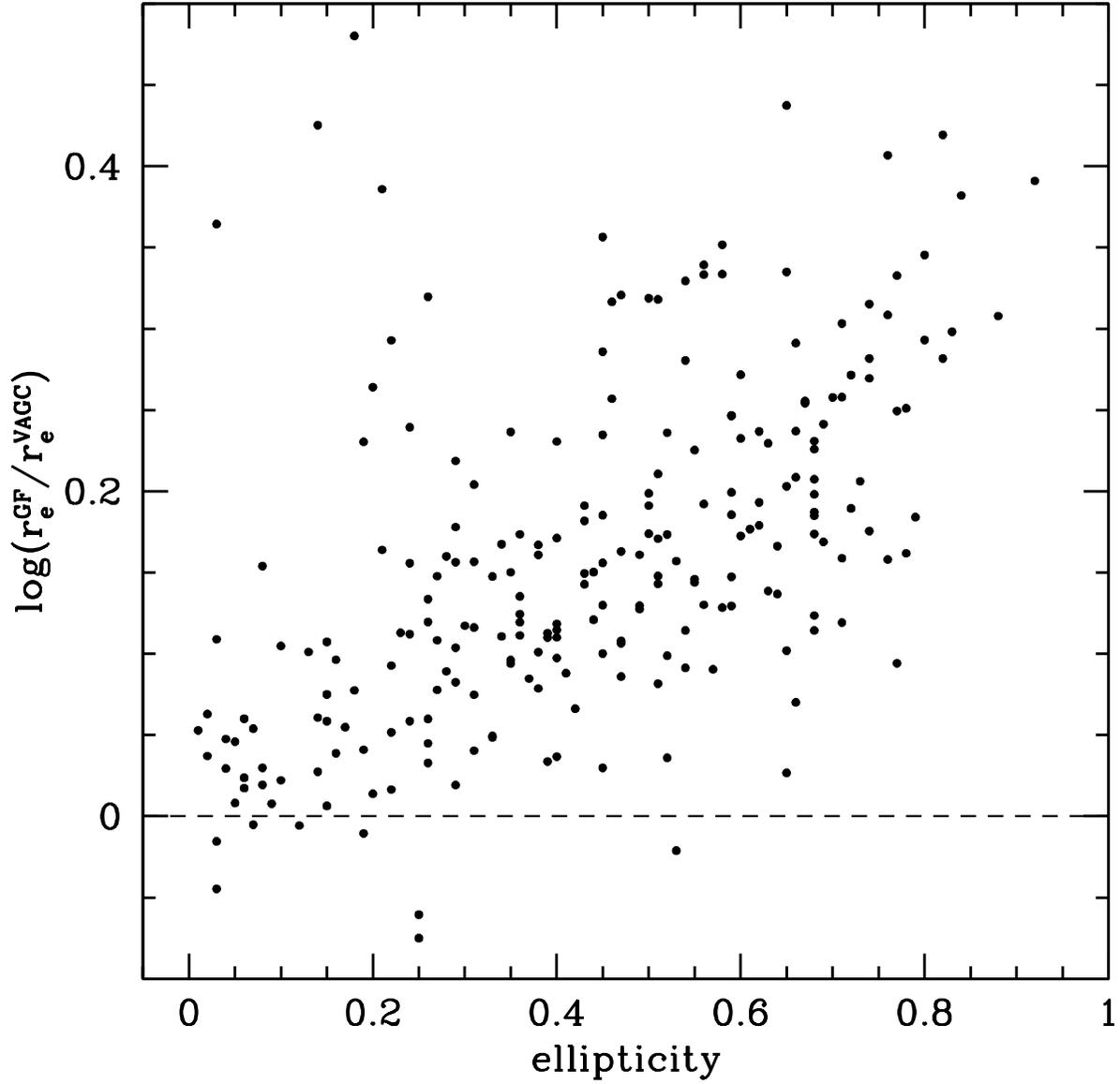}
\caption{The logarithm of the ratios of $r_e$ measured by GALFIT and VAGC
versus apparent ellipticity. The ellipticity stems from GALFIT.\label{ap2}}
\end{figure}

During the analysis of the structural properties of the two samples we realized
that there are systematic differences between the S\'ersic parameters provided
by the NYU VAGC and the results from GALFIT \citep{pen02} used for the GEMS
sample. The differences are larger than the expected errors and are most likely
caused by specifics of the two fitting methods applied. In order to confirm
this conjecture, we randomly selected 250 galaxies from SDSS covering the
entire luminosity range considered in this study. We fitted the surface
brightness distributions of this subsample with a S\'ersic model using GALFIT.
Five objects could not be fitted by GALFIT. The effective radii ($r_e$) of the
remaining 245 galaxies are compared to the $r_e$ provided by the NYU VAGC in
Figure \ref{ap1}. Obviously, GALFIT is systematically measuring larger radii.
The fitting method used for the NYU VAGC is described in \citet{bla105}. There
appear to be two major differences in the fitting procedure compared to GALFIT:
(1) all pixels are weighted equally in NYU VAGC, whereas in GALFIT each pixel
is weighted by the Poisson noise, (2) in the NYU VAGC a axissymmetric model is
assumed, whereas in GALFIT the ellipticity and the position angle are included
in the fitting process. The latter discrepancy probably makes the largest
contribution to the differences observed, as suggested by Figure \ref{ap2}. The
largest differences occur in the most inclined objects, although with a very
large scatter. After fitting both samples using the same method and weighting
scheme as in GEMS, the differences in $r_e$ practically disappear (see $\S$
\ref{fitp}).

The presented comparison does not intend to judge the quality or correctness of
the two approaches, but it shows that one has to be very cautious by combining
model parameters stemming from different fitting methods.

\clearpage

\end{document}